\documentclass[pra, nofootinbib,  twocolumn,  eqsecnum]{revtex4}
\usepackage{graphics}

\newcommand {\psig}{\Sigma_}
\newcommand {\pxi}{\Xi_}
\newcommand {\apsig}[1]{\langle \Sigma_{#1} \rangle}
\newcommand {\apxi}[1]{\langle \Xi_{#1} \rangle}
\newcommand {\apsigxi}[2]{\langle \Sigma_{#1} \Xi_{#2} \rangle}
\newcommand {\avecsig}{\langle \vec{\Sigma} \rangle}
\newcommand {\vecsig}{\vec{\Sigma}}
\newcommand {\avecxi}{\langle \vec{\Xi} \rangle}

\newcommand {\Tr}{{\mbox{Tr}}}

\begin{document}
\title{Mapping the Schr\"{o}dinger picture of open quantum dynamics}
\author{Thomas F. Jordan}
\email[email: ]{tjordan@d.umn.edu}
\affiliation{Physics Department, University of Minnesota, Duluth, Minnesota 55812}
\author{Anil Shaji}
\email[email: ]{shaji@physics.utexas.edu}
\affiliation{The University of Texas at Austin, Center for Statistical Mechanics, 1 University Station C1609, Austin TX 78712}
\author{E. C. G. Sudarshan}
\email[email: ]{sudarshan@physics.utexas.edu}
\affiliation{The University of Texas at Austin, Center for Statistical Mechanics, 1 University Station C1609, Austin TX 78712}  

\begin{abstract}
For systems described by finite matrices, an affine form is developed for the maps that describe evolution of density matrices for a quantum system that interacts with another. This is established directly from the Heisenberg picture. It separates elements that depend only on the dynamics from those that depend on the state of the two systems. While the equivalent linear map is generally not completely positive, the homogeneous part of the affine maps is, and is shown to be composed of multiplication operations that come simply from the Hamiltonian for the larger system. The inhomogeneous part is shown to be zero if and only if the map does not increase the trace of the square of any density matrix. Properties are worked out in detail for two-qubit examples. 
\end{abstract}


\maketitle

\section{Introduction}\label{one}
From the beginning, our understanding of quantum mechanics has involved both the Heisenberg picture \cite{heisenberg25a,heisenberg25aa} and the Schr\"{o}dinger picture \cite{schrodinger26a,schrodinger26aa} and the relation between them \cite{dirac26a,schrodinger26b}. Full understanding has been for a quantum system that is closed, which means there is no need to consider that it might interact with anything else. An open quantum system is a subsystem $S$ of a larger system and interacts with the subsystem $R$ that is the remainder, or rest of the larger system (which could be a reservoir). The evolution in $S$ is considered to be the result of unitary Hamiltonian evolution in the larger system of $S$ and $R$ combined. The Heisenberg picture for $S$ is clear. A matrix that represents a physical quantity for $S$ (an observable) is changed by the unitary transformation that changes every matrix that represents a physical quantity for the larger system. The Schr\"{o}dinger picture for $S$ has {\em not} been fully described. 

The state of $S$ is generally not a pure state, even when the state of the larger system of $S$ and $R$ combined is a pure state, so there is no Schr\"{o}dinger wave function and no Schr\"{o}dinger equation for $S$. There is a density matrix that describes the state of $S$. We can expect \cite{sudarshan61a,jordan61,jordan62,kraus71a,alicki87,breuer02,stelmachovic01a,sudarshan03a,jordan04,jordan04b} that evolution in the Schr\"{o}dinger picture for $S$ will be described by linear maps of matrices applied to the density matrix for $S$. Attention has been focused on the particular case where the initial state of the larger system is described by a density matrix that is a product of a density matrix for $S$ and a density matrix for $R$. Then the evolution in the Schr\"{o}dinger picture for $S$ is described by linear maps that are completely positive. They have been studied extensively \cite{sudarshan61a,kraus71a,choi72,choi74,choi75a,choi75b,kraus83,schumacher96a,chuang97}.

Recently we considered the general case, where $S$ and $R$ may be entangled in the initial state, so that the density matrix for the initial state of $S$ and $R$ combined is not a product. We worked out examples for entangled qubits in some detail, and for any system described by finite matrices we showed how the evolution in the Schr\"{o}dinger picture for $S$ can be described by linear maps \cite{jordan04}.

There are observations to be made of properties and structure in what has been found. They show us new features of the Schr\"{o}dinger picture for open quantum systems. Our first observation \cite{jordan04} was that the maps generally are not completely positive, and apply in limited domains \cite{pechukas94,stelmachovic01a}. Another observation \cite{jordan04b} is that putting the maps in an affine form \cite{chuang97,romero04a}, with homogeneous and inhomogeneous parts, can separate elements that depend only on the dynamics from those that depend on the state of entanglement. This gives a picture that is simpler in some respects and may be easier to use. We develop that picture here, show directly how it relates to the Heisenberg picture, and find some new properties of the affine form. Then we work out a new larger set of examples for two entangled qubits.

We do not consider equations of motion. A simple equation of motion would require that when the map that describes evolution for a time $t_1$ is followed by the map that describes evolution for a time $t_2$, the result is the map that describes evolution for the time $t_1+t_2$. The maps for open quantum systems generally do not have this semi-group property. To assume that they do is to make an approximation \cite{rau62a,gorini76a,lindblad76,davies76,lidar01a,sudarshan03b}. We want to simply describe the Schr\"{o}dinger picture before considering approximations to it. 

\section{Framework}\label{two}

We consider two interacting quantum systems $S$ and $R$, both described by finite  matrices: $N$$\times$$N$ matrices for $S$ and $M$$\times$$M$ for $R$. We use the matrices $F_{\mu 0}$, $F_{0 \nu}$ and $F_{\mu \nu}$ described in our previous paper \cite{jordan04}. The $F_{\mu 0}$ for $\mu = 0,1, \ldots N^2-1$ are $N^2$ Hermitian matrices for $S$ such that $F_{00}$ is $\openone_S$, the unit matrix for $S$, and
\begin{equation}
  \label{eq:f1}
  \Tr_S \left[ F_{\mu 0} F_{\nu 0} \right] = N \delta_{\mu \nu}.
\end{equation}
This implies that the $F_{\mu 0}$ are linearly independent, so every matrix for $S$ is a linear combination of the $F_{\mu 0}$. For example, the $F_{\mu 0}$ for $\mu = 1,2, \ldots N^2-1$ could be obtained by normalizing standard generators \cite{tilma02a} of $SU(N)$. The $F_{0 \nu}$ for $\nu = 0, 1, \ldots M^2-1$ are $M^2$ Hermitian matrices for $R$ such that $F_{00}$ is $\openone_R$, the unit matrix for $R$, and
\begin{equation}
  \label{eq:f2}
  \Tr_R \left[ F_{0 \mu} F_{0 \nu} \right] = M \delta_{\mu \nu}.
\end{equation}
Every matrix for $R$ is a linear combination of the $F_{0 \nu}$. We use notation that identifies $F_{\mu 0}$ with $F_{\mu 0} \otimes \openone_R$ and $F_{0 \nu}$ with $\openone_S \otimes F_{0 \nu}$ and let
\begin{equation}
  \label{eq:f3}
  F_{\mu \nu} = F_{\mu 0} \otimes F_{0 \nu}.
\end{equation}
Every matrix for the system of $S$ and $R$ combined is a linear combination of the $F_{\mu \nu}$.

We follow common physics practice and write a product of operators for separate systems, for example a product of Pauli matrices $\Sigma$ and $\Xi$ for the two qubits considered in Section \ref{seven}, simply as $\Sigma \Xi$, not $\Sigma \otimes \Xi$. Occasionally we insert a $\otimes$ for emphasis or clarity. 

The matrices $F_{\mu 0}$ for $\mu=1,2,\ldots N^2-1$ and $F_{0 \nu}$ for $\nu=1,2,\ldots M^2-1$ are generalizations of Pauli matrices (and like the Pauli matrices they have zero trace). We use them to describe density matrices the way we use Pauli matrices to describe density matrices for qubits. If $\Pi$ is a density matrix for the system of $S$ and $R$ combined, then
\begin{equation}
  \label{eq:f4}
  \Pi = \frac{1}{NM} \left( \openone + \sum_{\alpha =1}^{N^2-1} \langle F_{\alpha 0} \rangle F_{\alpha 0} + \sum_{\alpha=0}^{N^2-1}\sum_{\beta =1}^{M^2-1} \langle F_{\alpha \beta} \rangle F_{\alpha \beta} \right)
\end{equation}
and the density matrix $\rho$ for the subsystem $S$ is
\begin{equation}
  \label{eq:f5}
  \rho = \Tr_R \Pi = \frac{1}{N} \left( \openone + \sum_{\alpha=1}^{N^2-1} \langle F_{\alpha 0} \rangle F_{\alpha 0} \right)
\end{equation}
so that
\begin{equation}
  \label{eq:f6}
  \langle F_{\alpha \beta} \rangle = \Tr\left[ F_{\alpha \beta} \Pi \right]
\end{equation}
and in particular
\begin{equation}
  \label{eq:f7}
  \langle F_{\alpha 0} \rangle = \Tr_S \left[ F_{\alpha 0} \, \Tr_R \Pi \right] = \Tr_S \left[ F_{\alpha 0} \rho \right].
\end{equation}

If $U$ is a unitary matrix, then
\begin{equation}
  \label{eq:f8}
  U^{\dagger}F_{\mu \nu} U = \sum_{\alpha =0}^{N^2-1} \sum_{\beta =0}^{M^2-1}t_{\mu \nu \,;\, \alpha \beta}F_{\alpha \beta}
\end{equation}
with the $t_{\mu \nu \,;\, \alpha \beta}$ elements of a real orthogonal matrix, so $t^{-1}_{\mu \nu \,;\, \alpha \beta}$ is $t_{\alpha \beta \,;\, \mu \nu}$. Since $U^{\dagger} \openone U$ and $U \openone U^{\dagger}$ are $\openone$,
\begin{equation}
  \label{eq:f9}
  t_{00 \,;\, \alpha \beta} = \delta_{0\alpha} \, \delta_{0 \beta} \;, \quad t_{\alpha \beta \,;\, 00} = \delta_{\alpha 0} \, \delta_{\beta 0}.
\end{equation}

\section{Affine maps of density matrices}\label{three}

Suppose that in the system of $S$ and $R$ combined the matrices $C$ that represent physical quantities are changed to $U^{\dagger}CU$ by a unitary operator $U$. This is the Heisenberg picture. The mean values are changed to
\begin{equation}
  \label{eq:a1}
  \langle C \rangle^U = \langle U^{\dagger}CU \rangle = \Tr \left[U^{\dagger}C U \; \Pi \right] = \Tr \left[ C \;U \Pi U^{\dagger} \right].
\end{equation}
The result is the same if the matrices $C$ are left unchanged and the density matrix $\Pi$ is changed to $U \Pi U^{\dagger}$. This is the Schr\"{o}dinger picture. 

Let $A$ be a matrix for the subsystem $S$. In the Heisenberg picture it is changed to $U^{\dagger}AU$ so its mean value is changed to
\begin{equation}
  \label{eq:a2}
  \langle A \rangle^U = \langle U^{\dagger}A U \rangle = \Tr_S \left[ A \; \Tr_R \left[U \Pi U^{\dagger} \right] \right].
\end{equation}
The Schr\"{o}dinger picture for the subsystem $S$ is that the density matrix $\rho$ for $S$ is changed to
\begin{equation}
  \label{eq:a3}
  \rho^U = \Tr_R \left[ U \Pi U^{\dagger} \right] = L(\rho) + K
\end{equation}
where
\begin{eqnarray}
  \label{eq:a4}
  L(Q) & = & \Tr_R \left[ U \, Q \otimes \frac{\openone}{M} U^{\dagger} \right] \nonumber \\
  K & = & \Tr_R \left[ U \left(\Pi - \rho \otimes \frac{\openone}{M} \right)U^{\dagger} \right].
\end{eqnarray}
The $L$ is a completely positive linear map that applies to any matrix $Q$ for the subsystem $S$, density matrix or not. It has the property that $L(\openone)$ is $\openone$. The map $L$ depends on $U$ but does not depend on the state of $R$ or on the state of entanglement of the subsystems $S$ and $R$. 

The $K$ is the only part of $\rho^U$ that can depend on the state of $R$ or on the correlations between $S$ and $R$. With the same $K$, the Eq. (\ref{eq:a3}) defines a map that applies to different density matrices $\rho$ representing different states of $S$. The state of $S$ can be changed without changing $K$. That is evident from Eq. (\ref{eq:a4}). Since $\Pi$ and $\rho \otimes \frac{\openone}{M}$ are density matrices that give the same mean values for any matrix $A$ for $S$,
\begin{equation}
  \label{eq:a5}
  \Tr \left[ A \Pi \right] = \Tr_S \left[ A \, \Tr_R \Pi  \right] = \Tr_S \left[ A \rho \right] = \Tr \left[ A \; \rho \otimes \frac{\openone}{M} \right],
\end{equation}
their difference does not need to change when the state of $S$ is changed. Explicitly, from Eqs. (\ref{eq:f4}) and (\ref{eq:f5}) 
we see that
\begin{equation}
  \label{eq:a8}
  \Pi - \rho \otimes \frac{\openone}{M} = \frac{1}{NM} \sum_{\alpha =0}^{N^2-1} \sum_{\beta=1}^{M^2-1} \langle F_{\alpha \beta} \rangle F_{\alpha \beta},
\end{equation}
which does not depend on the $\langle F_{\alpha 0} \rangle$ which describe the state of $S$. 

When we define a map, we consider all the $\langle F_{\alpha \beta} \rangle$ to be independent. The $\langle F_{\alpha 0} \rangle$ describe the state of $S$. The $\langle F_{\alpha \beta} \rangle$ for $\beta$ not $0$ are considered to be parameters of the map that describe the effect of the dynamics of the larger system of $S$ and $R$ combined that drives the evolution of $S$. Different $\langle F_{\alpha \beta} \rangle$ for $\beta$ not $0$ specify different maps. Each map applies to different states of $S$ described by different $\langle F_{\alpha 0} \rangle$. For each map there is one $N$$\times$$N$ matrix $K$. We explained this with examples in our previous paper \cite{jordan04}. We also mentioned there that an alternative map can be used in the special case of a product state where $\langle F_{\alpha \beta} \rangle$ is $\langle F_{\alpha 0} \rangle \langle F_{0 \beta} \rangle$; we will not consider that here.
 
In the Schr\"{o}dinger picture $K$ accounts for the parts of mean values $\langle A \rangle^U$ that in the Heisenberg picture come from matrices $U^{\dagger}AU$ not being matrices for $S$. Without $K$, a mean value $\langle A \rangle^U$ calculated in the Schr\"{o}dinger picture would be
\begin{eqnarray}
  \label{eq:a9}
  \langle A \rangle_0^U &=& \Tr_S \left[ A L (\rho) \right]  = \Tr_S \left[ A \; \Tr_R\left[ U \rho \otimes \frac{\openone}{M} U^{\dagger} \right] \right] \nonumber 
  \\ &=& \Tr \left[U^{\dagger}AU \; \rho \otimes \frac{\openone}{M} \right]
\end{eqnarray}
which is obtained in the Heisenberg picture by replacing $\Pi$ with $\rho \otimes \frac{\openone}{M}$, which cuts off the part of $U^{\dagger}AU$ that is not a matrix for $S$. The full mean value $\langle A \rangle^U$ is obtained by adding
\begin{equation}
 \label{eq:a10}
  \Tr_S \left[ A K \right] = \Tr \left[ U^{\dagger}AU \; \left( \Pi - \rho \otimes \frac{\openone}{M} \right) \right].
\end{equation}
This equation (\ref{eq:a10}) follows directly from Eq. (\ref{eq:a4}). In particular, we have
\begin{equation}
  \label{eq:a11}
  \Tr_S \left[ F_{\mu 0} K \right] = \Tr \left[ U^{\dagger} F_{\mu 0} U \; \left( \Pi - \rho \otimes \frac{\openone}{M} \right) \right]
\end{equation}
for $\mu = 1, 2, \ldots N^2-1$, so, because $\Tr_S K$ is zero,
\begin{eqnarray}
  \label{eq:a12}
  K &=& \sum_{\mu=1}^{N^2-1} \Tr_S \left[ F_{\mu 0} K \right] \, F_{\mu 0} \nonumber 
  \\ &=& \sum_{\mu =1}^{N^2-1} \Tr \left[ U^{\dagger} F_{\mu 0} U \; \left( \Pi - \rho \otimes \frac{\openone}{M} \right) \right]\, F_{\mu 0}. \nonumber \\
\end{eqnarray}
This is how we actually calculate $K$, as in the examples for two qubits described in Section \ref{seven}. We do not need to calculate $U\Pi U^{\dagger}$ for the whole system of $S$ and $R$ combined. We just calculate $U^{\dagger}F_{\mu 0} U$ for the basis matrices $F_{\mu 0}$ for $S$, take the mean values of the parts that extend outside the matrices for $S$, and get $K$ from Eq. (\ref{eq:a12}).

\section{Purity decrease} \label{four}

A property that depends simply on the presence or absence of $K$ is that 
\begin{equation}
  \label{eq:p1}
  \Tr \left[ \left(\rho^U \right)^2 \right] \leq \Tr \left[ \rho^2 \right]
\end{equation}
for all density matrices $\rho$ if and only if $K$ is zero. Here is a proof. If $K$ is zero then from Eqs. (\ref{eq:f5}), (\ref{eq:a9}) and (\ref{eq:f8}) 
\begin{equation}
  \label{eq:p3}
  \rho^U = L(\rho) = \frac{1}{N} \left( \openone + \sum_{\mu=1}^{N^2-1} \langle F_{\mu 0} \rangle_0^U F_{\mu 0} \right)
\end{equation}
where
\begin{equation}
  \label{eq:p4}
  \langle F_{\mu 0} \rangle_0^U = \Tr \left[ U^{\dagger} F_{\mu 0} U \; \rho \otimes \frac{\openone}{M} \right] = \sum_{\alpha=1}^{N^2-1} t_{\mu 0 \,;\, \alpha 0 } \langle F_{\alpha 0} \rangle
\end{equation}
because $\Tr \left[ F_{\alpha \beta} \, \rho \otimes \frac{\openone}{M} \right]$ is zero if $\beta$ is not zero and $t_{\mu 0 \, ;\, 00}$ is zero when $\mu$ is not zero. Let
\begin{equation}
  \label{eq:p5}
  \chi_{\alpha \beta} = \langle F_{\alpha 0} \rangle \quad {\mbox{for}} \quad \alpha =1,2,\ldots N^2-1\,, \quad \beta=0
\end{equation}
and $\chi_{\alpha \beta}=0$ for other $\alpha$, $\beta$, and let
\begin{equation}
  \label{eq:p6}
  \chi'_{\mu \nu} = \sum_{\alpha=0}^{N^2-1} \sum_{\beta=0}^{M^2-1} t_{\mu \nu \,;\, \alpha \beta} \chi_{\alpha \beta}.
\end{equation}
Then $\langle F_{\mu 0} \rangle_0^U$ is $\chi'_{\mu 0}$ and
\begin{eqnarray}
  \label{eq:p7}
  \sum_{\mu=1}^{N^2-1} \left( \langle F_{\mu 0}\rangle_0^U \right)^2 & \leq &\sum_{\mu=0}^{N^2-1} \sum_{\nu=0}^{M^2-1} \left( \chi'_{\mu \nu} \right)^2 \nonumber 
  \\ &=& \sum_{\alpha=0}^{N^2-1} \sum_{\beta=0}^{M^2-1} \left(\chi_{\alpha \beta} \right)^2 \nonumber 
  \\  &=& \sum_{\alpha=1}^{N^2-1} \langle F_{\alpha 0} \rangle^2
\end{eqnarray}
which implies the inequality (\ref{eq:p1}).

Suppose $K$ is not zero. Then (\ref{eq:p1}) fails for at least one density matrix $\rho$. Let
\begin{equation}
  \label{eq:p8}
  K = \sum_{n=1}^N \lambda_n |n \rangle \langle n |
\end{equation}
\begin{equation}
  \label{eq:p9}
  \rho = \frac{\openone}{N} = \frac{1}{N} \sum_{n=1}^N |n \rangle \langle n |. 
\end{equation}
For this $\rho$ we have
\begin{equation}
  \label{eq:p10}
  \rho^U = L(\rho) + K = \frac{\openone}{N} +K = \sum_{n=1}^N \left( \frac{1}{N} + \lambda_n \right)|n \rangle \langle n |
\end{equation}
and, since
\begin{equation}
  \label{eq:p11}
  \sum_{n=1}^N \lambda_n = \Tr K = 0,
\end{equation}
\begin{eqnarray}
  \label{eq:p12}
  \Tr \left[ \left( \rho^U \right)^2 \right] &=& \sum_{n=1}^N \left( \frac{1}{N} + \lambda_n \right)^2 \nonumber 
  \\ &=& \sum_{n=1}^N \left(\frac{1}{N} \right)^2 + \sum_{n=1}^N \left( \lambda_n \right)^2 > \sum_{n=1}^N \left(\frac{1}{N} \right)^2 \nonumber
 \\ & = & \Tr \left[ \rho^2 \right].
\end{eqnarray}

For this proof we assume that the inequality (\ref{eq:p1})holds when $\rho$ is $\openone/N$. A map generally is meant to apply only to a limited set of density matrices $\rho$, where it represents the result of the unitary Hamiltonian dynamics in the larger system of $S$ and $R$ combined. The examples worked out in Section \ref{seven} show there are maps that are not meant to apply when $\rho$ is $\openone/N$. In such cases, the assumption that the inequality (\ref{eq:p1}) holds when $\rho$ is  $\openone/N$ is a mathematical statement that does not have a direct physical interpretation. 

\section{Map operations} \label{five}
The map $L$ can be done with multiplication operations simply related to $U$. Let
\begin{equation}
  \label{eq:o1}
  U = \sum_{\nu=0}^{M^2-1}G(\nu)F_{0 \nu}
\end{equation}
with the $G(\nu)$ matrices for $S$. Then
\begin{equation}
  \label{eq:o2}
  L(Q) = \sum_{\nu =0}^{M^2-1}G(\nu) Q G(\nu)^{\dagger},
\end{equation}
\begin{equation}
  \label{eq:o3}
  \sum_{\nu=0}^{M^2-1}G(\nu)^{\dagger}G(\nu) = \frac{1}{M} \Tr_R \left[ U^{\dagger}U \right] = \openone_S 
\end{equation}
and
\begin{equation}
  \label{eq:o3a}
  L(\openone_S) = \sum_{\nu=0}^{M^2-1}G(\nu)G(\nu)^{\dagger} = \frac{1}{M} \Tr_R \left[ U U^{\dagger} \right] = \openone_S.
\end{equation}
Altogether
\begin{equation}
  \label{eq:o4}
  \rho^U = \sum_{\nu=0}^{M^2-1}G(\nu) \rho G(\nu)^{\dagger} + K.
\end{equation}
The matrices $G(\nu)$ depend on $U$ and depend on the choice of basis matrices $F_{0 \nu}$. Making that choice to conform with $U$ can simplify the set of matrices $G(\nu)$, as the examples described in Section \ref{seven} will show. The matrices $G(\nu)$ do not depend on the state of $S$ and $R$. They can be calculated from $U$ and used for any states. 

\section{Linear maps of matrices} \label{six}

We fill out the Schr\"{o}dinger picture with a linear map of matrices $Q$ for $S$ that gives $\rho^U$ when applied to a density matrix $\rho$. It is
\begin{equation}
  \label{eq:l1}
  Q \longrightarrow Q' = L(Q) + K \Tr Q
\end{equation}
or, in terms of the basis matrices,
\begin{equation}
  \label{eq:l2}
  \openone' = \openone + NK\;, \quad F_{\alpha 0}' = L(F_{\alpha 0})
\end{equation}
for $\alpha=1,2,\ldots N^2-1$. This is the only linear map that can give $\rho^U$ for a variety of density matrices $\rho$ described by Eq. (\ref{eq:f5}). Since $K$ is the same for all $\rho$, it cannot come from the terms with variable coefficients $\langle F_{\alpha 0} \rangle$. It can only be part of $\openone'$.

We described this map in our previous paper \cite{jordan04}. We approached it differently there. We considered first the map of mean values $\langle F_{\alpha 0} \rangle$ for the basis matrices for $S$ and then the consequent maps of density matrices $\rho$ and of the basis matrices $\openone$ and $F_{\alpha 0}$. By working out examples of two entangled qubits, we found that this linear map (\ref{eq:l1}) is generally not completely positive, that there is a limited domain in which it maps every positive matrix to a positive matrix, and that there is a limited domain in which it represents the effect of the dynamics of the larger system. We call these domains the positivity domain and the compatibility domain. We considered the description of the linear map (\ref{eq:l1}) by
\begin{equation}
  \label{eq:l4}
  Q'_{rs} = \sum_{j,k=1}^N B_{rj \,;\, sk} Q_{jk}
\end{equation}
and by
\begin{equation}
  \label{eq:l5}
  Q' = \sum_{n=1}^p C(n) Q C(n)^{\dagger} - \sum_{n=p+1}^{N^2} C(n) Q C(n)^{\dagger}.
\end{equation}

Here we only make two comments. From Eq. (\ref{eq:l1}) we see that $K$ contributes $K_{rs}  \, \delta_{jk}$ to $B_{rj \,;\, sk}$ in Eq. (\ref{eq:l4}). From Eqs (\ref{eq:l1}) and (\ref{eq:o2}) we have
\begin{equation}
  \label{eq:l6}
  Q' = \sum_{\nu=0}^{M^2-1} G(\nu) Q G(\nu)^{\dagger} + K \Tr Q.
\end{equation}
In our experience with examples, the matrices $G(\nu)$ and $K$ in this equation have been significantly simpler than the matrices $C(n)$ in Eq. (\ref{eq:l5}). 

\section{Quantum process tomography} \label{sixa}

How is such a map found? Is it observable? What can be seen in experiments? Is the map determined \cite{chuang97} by the effect of the dynamics on different density matrices $\rho$? It is if the compatibility domain contains an open set of values for the $\langle F_{\alpha 0} \rangle$. Then for each $\alpha$ from $1$ to $N^2 -1$ there are states of $S$, with density matrices $\rho$ described by Eq. (\ref{eq:f5}), that differ only in the value of $\langle F_{\alpha 0} \rangle$ for that one $\alpha$. Between two of these states, the difference in
\begin{equation}
  \label{eq:tom1}
  \rho^U = \frac{1}{N} \left( \openone' + \sum_{\alpha =1}^{N^2-1} \langle F_{\alpha 0} \rangle F_{\alpha 0}' \right)
\end{equation}
is just
\begin{equation}
  \label{eq:tom2}
  \Delta \rho^U = \frac{1}{N} \left( \Delta \langle F_{\alpha 0} \rangle \right) F_{\alpha 0}'.
\end{equation}
This determines $F_{\alpha 0}'$. The map is specified by $\openone'$ and these $F_{\alpha 0}'$ for $\alpha$ from $1$ to $N^2-1$. When all the $F_{\alpha 0}'$ are known, $\openone'$ can be found from any $\rho^U$. If the compatibility domain contains the state where $\langle F_{\alpha 0} \rangle$ is zero for all $\alpha$ from $1$ to $N^2-1$, so $\rho$ is $\frac{1}{N} \openone$, then $\openone'$ is determined by
\begin{equation}
  \label{eq:tom3}
  \rho^U = \frac{1}{N} \openone'
\end{equation}
for that state, but as examples described in Section \ref{seven} show, the compatibility domain does not always include that state. 

The compatibility domain is the set of density matrices $\rho$ that can be affected by the dynamics for the states being considered for the larger system of $S$ and $R$ combined. If there are enough density matrices in the compatibility domain that are accessible to experiments, the map can be determined experimentally. The choice of the density matrices $\rho$ to be used \cite{chuang97} will depend on the particular situation. Density matrices that are handy for one situation may not even be in the compatibility domain for another situation.

\section{Two-qubit examples} \label{seven}

We consider two qubits described by Pauli matrices $\psig 1$, $\psig 2$, $\psig 3$ for $S$ and $\pxi 1$, $\pxi 2$, $\pxi 3$ for $R$, so $F_{j0}$ is $\psig j$ and $F_{0k}$ is $\pxi k$, which implies $F_{jk}$ is $\psig j \pxi k$, for $j,k=1,2,3$. The density matrix for the two qubits is
\begin{eqnarray}
  \label{eq:tq1}
  \Pi & = & \frac{1}{4} \left( \openone + \sum_{j=1}^3 \apsig{j} \psig j + \sum_{k=1}^3 \apxi{k} \pxi k \right. \nonumber 
  \\ && \hspace{2.3 cm} \left. + \sum_{j,k=1}^3 \apsigxi{j}{k} \psig j \pxi k \right)
\end{eqnarray}
and the density matrix for $S$ is
\begin{equation}
  \label{eq:tq2}
  \rho = \Tr_R \Pi = \frac{1}{2} \left( \openone + \sum_{j=1}^3 \apsig{j} \psig j \right).
\end{equation}
We let
\begin{equation}
  \label{eq:tq3}
 K = \frac{1}{2} \sum_{j=1}^3 \kappa_j \psig j
\end{equation}
so
\begin{equation}
  \label{eq:tq4}
  \kappa_j = \Tr_S \left[ \psig j K \right].
\end{equation}
We write $\avecsig$ for the vector with components $\apsig{1}$, $\apsig{2}$, $\apsig{3}$ and $\vec{\kappa}$ for the vector with components $\kappa_1$, $\kappa_2$, $\kappa_3$, and write $|\avecsig|$ and $| \vec{\kappa} |$ for the lengths of these vectors. We write $\vecsig$ for the vector whose components are the matrices $\psig 1$, $\psig 2$, $\psig 3$.

Our examples are for different unitary matrices $U$. For the first set we think of $U$ as describing the dynamics of the two qubits. In the second set $U$ describes a Lorentz transformation of the spin of a massive particle for states with two possible values of the momentum. This illustrates how the maps developed for dynamics can be used for other transformations as well. 

\subsection{Interaction Hamiltonians}\label{interaction}

The first examples are motivated by considering Hamiltonians that have only interaction terms, no free Hamiltonian terms, as in an interaction picture. From $\sum_{j,k=1}^3 \gamma'_{jk} \psig j \pxi k$ we can get $\sum_{j=1}^3 \gamma_j \psig j \pxi j$ by making a rotation in each qubit \cite{zhang03a} and redefining the $\psig j$ and $\pxi k$, so to choose an example we let
\begin{equation}
  \label{eq:int1}
  U = e^{-i \frac{1}{2} \left( \gamma_1 \psig 1 \pxi 1 + \gamma_2 \psig 2 \pxi 2 + \gamma_3 \psig 3 \pxi 3 \right)}
\end{equation}
(where $\gamma_1$, $\gamma_2$, $\gamma_3$ can be functions of time). The three matrices $\psig 1 \pxi 1$, $\psig 2 \pxi 2$, $\psig 3 \pxi 3$ commute with each other (The different $\psig j$ anticommute and the different $\pxi j$ anticommute, so the different $\psig j \pxi j$ commute). That allows us to easily compute
\begin{eqnarray}
  \label{eq:int2}
  U^{\dagger}\psig 1 U & = & \psig 1 e^{-i \gamma_2 \psig 2 \pxi 2} e^{-i \gamma_3 \psig 3 \pxi 3} \nonumber \\
 &=& \psig 1 \cos \gamma_2 \cos \gamma_3 + \pxi 1 \sin \gamma_2 \sin \gamma_3 \nonumber \\
 && \hspace{3 mm} - \psig 2 \pxi 3 \cos \gamma_2 \sin \gamma_3 + \psig 3 \pxi 2 \sin \gamma_2 \cos \gamma_3 \nonumber \\
\end{eqnarray}
using the algebra of Pauli matrices, and similarly
\begin{eqnarray}
  \label{eq:int3}
  U^{\dagger}\psig 2 U &=& \psig 2 \cos \gamma_3 \cos \gamma_1 + \pxi 2 \sin \gamma_3 \sin \gamma_1 \nonumber \\
 && \hspace{3 mm} - \psig 3 \pxi 1 \cos \gamma_3 \sin \gamma_1 + \psig 1 \pxi 3 \sin \gamma_3 \cos \gamma_1,  \nonumber \\
\end{eqnarray}
\begin{eqnarray}
  \label{eq:int4}
  U^{\dagger}\psig 3 U  &=& \psig 3 \cos \gamma_1 \cos \gamma_2 + \pxi 3 \sin \gamma_1 \sin \gamma_2 \nonumber \\
 && \hspace{3 mm} - \psig 1 \pxi 2 \cos \gamma_1 \sin \gamma_2 + \psig 2 \pxi 1 \sin \gamma_1 \cos \gamma_2.  \nonumber \\ 
\end{eqnarray}
Interchanging $U$ and $U^{\dagger}$ has the same effect as changing the sign of every $\gamma_j$. Thus we see that 
\begin{eqnarray}
  \label{eq:int5}
  L(\psig 1) & = & \psig 1 \cos \gamma_2 \cos \gamma_3 \nonumber \\
   L(\psig 2) & = & \psig 2 \cos \gamma_3 \cos \gamma_1 \nonumber \\
    L(\psig 3) & = & \psig 3 \cos \gamma_1 \cos \gamma_2.
\end{eqnarray}
Taking mean values in Eqs. (\ref{eq:int2}), (\ref{eq:int3}) and (\ref{eq:int4}) gives the $\apsig{j}^U$, from which we see that
\begin{eqnarray}
  \label{eq:int6}
  \kappa_1 & = & \apxi{1} \sin \gamma_2 \sin \gamma_3 - \apsigxi{2}{3} \cos \gamma_2 \sin \gamma_3 \nonumber 
  \\ && \hspace{3.5 cm} + \apsigxi{3}{2} \sin \gamma_2 \cos \gamma_3 \nonumber \\ \nonumber \\
  \kappa_2 & = & \apxi{2} \sin \gamma_3 \sin \gamma_1 - \apsigxi{3}{1} \cos \gamma_3 \sin \gamma_1
  \nonumber 
  \\ && \hspace{3.5 cm} + \apsigxi{1}{3} \sin \gamma_3 \cos \gamma_1 \nonumber \\ 
  \kappa_3 & = & \apxi{3} \sin \gamma_1 \sin \gamma_2 - \apsigxi{1}{2} \cos \gamma_1 \sin \gamma_2 
  \nonumber 
  \\ && \hspace{3.5 cm} + \apsigxi{2}{1} \sin \gamma_1 \cos \gamma_2. \nonumber \\
\end{eqnarray}
We can construct $K$ from Eq. (\ref{eq:tq3}) and then get the linear map from Eq. (\ref{eq:l1}) or use Eq. (\ref{eq:l2}) to get the 
\begin{equation}
  \label{eq:int7}
  \psig j' = L (\psig j) \,, \quad \openone' = 1 + 2 K 
\end{equation}
which determine the linear map. For the description of the linear map by Eq. (\ref{eq:l4}) we find that
\begin{widetext}
\begin{equation}
  \label{eq:int8}
  B = \frac{1}{2} \left( \begin{array}{cccc} 1+\kappa_3+ C_1 C_2 & 0 & \kappa_1 - i\kappa_2 &C_2 C_3 +  C_3C_1 \\
0 & 1+\kappa_3 - C_1 C_2 &  C_2 C_3 -  C_3 C_1 & \kappa_1 - i\kappa_2 \\
\kappa_1 + i \kappa_2 & C_2 C_3 - C_3 C_1 & 1 - \kappa_3 -C_1 C_2 & 0 \\
C_2 C_3 +  C_3 C_1 & \kappa_1 + i \kappa_2 & 0 &  1 - \kappa_3 + C_1 C_2
 \end{array} \right)
\end{equation}
where $C_i \equiv \cos \gamma_i$ for $i=1,2,3$, and the rows and columns of the matrix are in the order $11$, $12$, $21$, $22$. You can check that this is correct because it does give $\psig 1'$, $\psig 2'$, $\psig 3'$, $\openone'$ that agree with Eq. (\ref{eq:int7}).
\begin{figure}[!h]
   \begin{center}
   \includegraphics{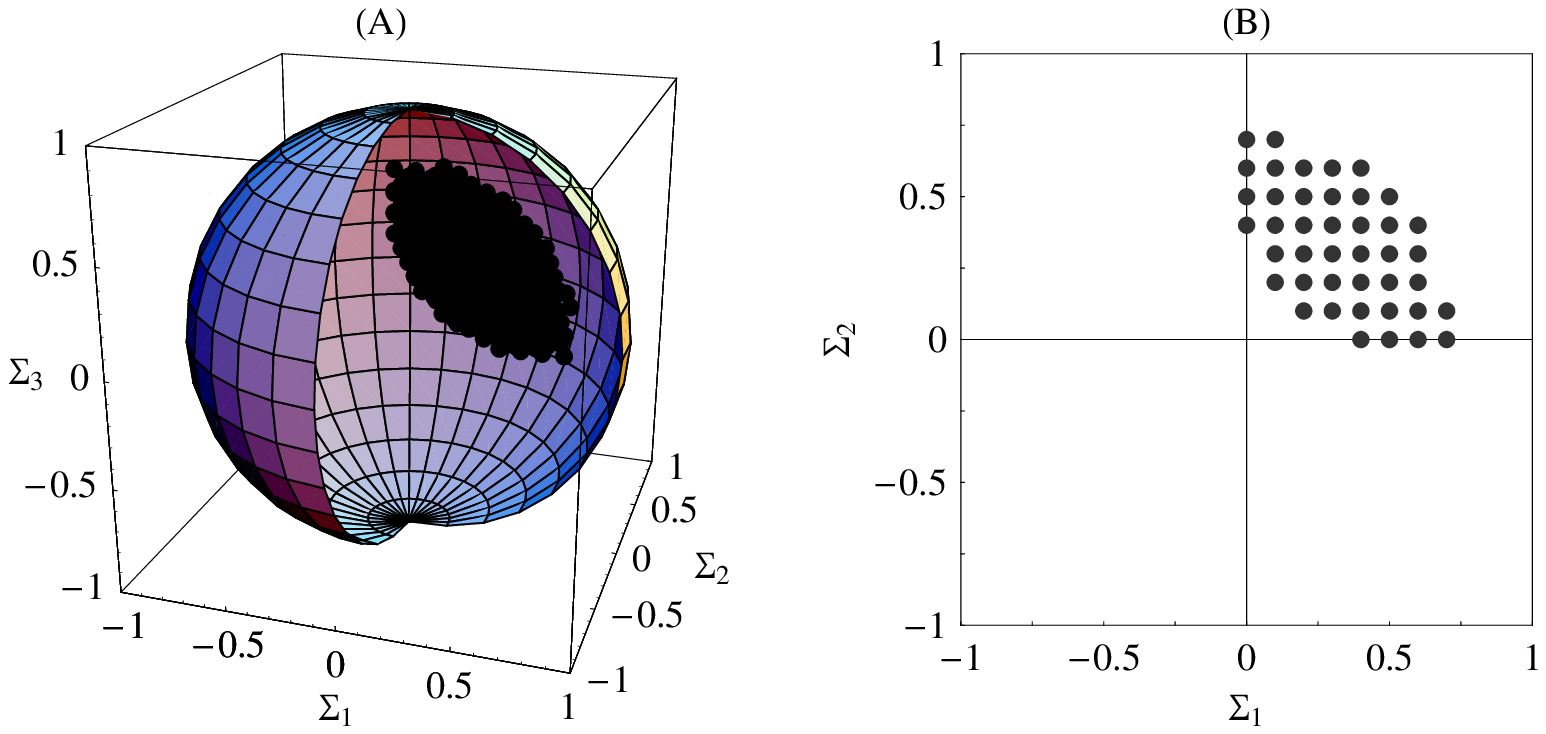}
   \end{center}
   \caption{(color online) The compatibility domain for $\apxi{1}$, $\apxi{2}$, $\apxi{3}$, $\apsigxi{1}{2}$, $\apsigxi{1}{3}$, $\apsigxi{2}{1}$, $\apsigxi{2}{3}$, $\apsigxi{3}{1}$ and $\apsigxi{3}{2}$ all equal to $\frac{1}{4}$. (A) The whole compatibility domain inside the unit sphere. (B) The section of the domain in the $(\psig 1 , \psig 2)$ plane. The sections of the domain in the $(\psig 1 , \psig 3)$ and $(\psig 2 , \psig 3)$ planes are identical.}
\label{fig1}
\end{figure}

\end{widetext}

The examples described in our previous paper \cite{jordan04} are obtained as a particular case by letting $\gamma_1$ and $\gamma_2$ be zero, taking $\gamma_3$ to be $\omega t$, and changing the $\pxi 3$ in $U$ to $\pxi 1$. The new set of examples is much richer. In $U$ there are three parameters instead of one. In $K$ there are nine mean values: the three $\apxi{k}$ and the six $\apsigxi{j}{k}$ for $j \neq k$. The maps described in our previous paper depend only on $\apsigxi{1}{1}$ and $\apsigxi{2}{1}$.

Different values of the $\apxi{k}$ or $\apsigxi{j}{k}$ in $K$ generally give different maps. Each map is made to be used for a particular set of states described by a particular set of density matrices $\rho$, or a particular set of $\avecsig$, which we call the compatibility domain. It is the set of $\avecsig$ that are compatible with the $\apxi{k}$ and $\apsigxi{j}{k}$ in $K$ in describing a possible initial state for the two qubits. The increased number of $\apxi{k}$ and $\apsigxi{j}{k}$ in $K$ means that the compatibility domains are more restricted and varied. It is difficult to describe general features of the compatibility domains beyond the fact that they are convex \cite{jordan04}.

In a larger domain, which we call the positivity domain, the map takes every positive matrix to a positive matrix. The positivity domain is the set of $\avecsig$ for which $|\avecsig^U| \leq 1$. It depends on both the $\gamma_j$ in $U$ and the $\apxi{k}$ and $\apsigxi{j}{k}$ in $K$, so the variety of positivity domains is larger than the large variety of compatibility domains. We have looked at several examples.   

These examples exhibit new features. For the examples described previously \cite{jordan04}, the compatibility domain is not changed by reflection through the origin in the space of the $\avecsig$; if $\avecsig$ is in the compatibility domain, then so is $-\avecsig$. The origin, the zero $\avecsig$, is always in the compatibility domain. We can see from Figs. \ref{fig1}, \ref{fig1a} and \ref{fig2} that these properties do not hold as a general rule. Another property of the examples described previously \cite{jordan04} is that the compatibility domain is the intersection of all the positivity domains for the same values of the $\apxi{k}$ and $\apsigxi{j}{k}$ in $K$. We can easily see that this also is not generally true. There are simple cases where the zero $\avecsig$ is in every positivity domain but not in the compatibility domain. 

For example, suppose $\apxi{1}$ and $\apsigxi{3}{1}$ are positive and all the other $\apxi{k}$ and $\apsigxi{j}{k}$ for $j \neq k$ are zero. Then
\begin{eqnarray}
  \label{eq:int9}
  \kappa_1 &=& \apxi{1} \sin \gamma_2 \sin \gamma_3, \nonumber \\
  \kappa_2 &=& -\apsigxi{3}{1} \cos \gamma_3 \sin \gamma_1, \nonumber \\
  \kappa_3 &=& 0
\end{eqnarray}
and
\begin{equation}
  \label{eq:int10}
  |\vec{\kappa}|^2 \leq \sin^2 \gamma_3 + \cos^2 \gamma_3 =1.
\end{equation}

\noindent This implies that the zero $\avecsig$ is in all the positivity domains for different $\gamma_1$, $\gamma_2$, $\gamma_3$, because $\avecsig^U$ is $\vec{\kappa}$ when $\avecsig$ is zero.

We can see from Fig. \ref{fig2} that for cases of this kind there are compatibility domains that do not contain the zero $\avecsig$. We can easily show that there are many such cases. First we show that there is a substantial compatibility domain for any values of $\apxi{1}$ and $\apsigxi{3}{1}$ short of the limit where $\apxi{1}$ or $\apsigxi{3}{1}$ is 1. We find a set of $\avecsig$ for which the matrix
\begin{eqnarray}
  \label{eq:int11}
  \Pi &=& \frac{1}{4} \left( \openone + \apsig{1} \psig 1 + \apsig{2} \psig 2 + \apsig{3} \psig 3 \right.  \nonumber 
  \\ && \hspace{1.5 cm} \left. + \apsigxi{3}{1} \psig 3 \pxi 1 + \apxi{1} \pxi 1 \right)
\end{eqnarray}

\begin{widetext}

\begin{figure}[!h]
   \begin{center}
   \includegraphics{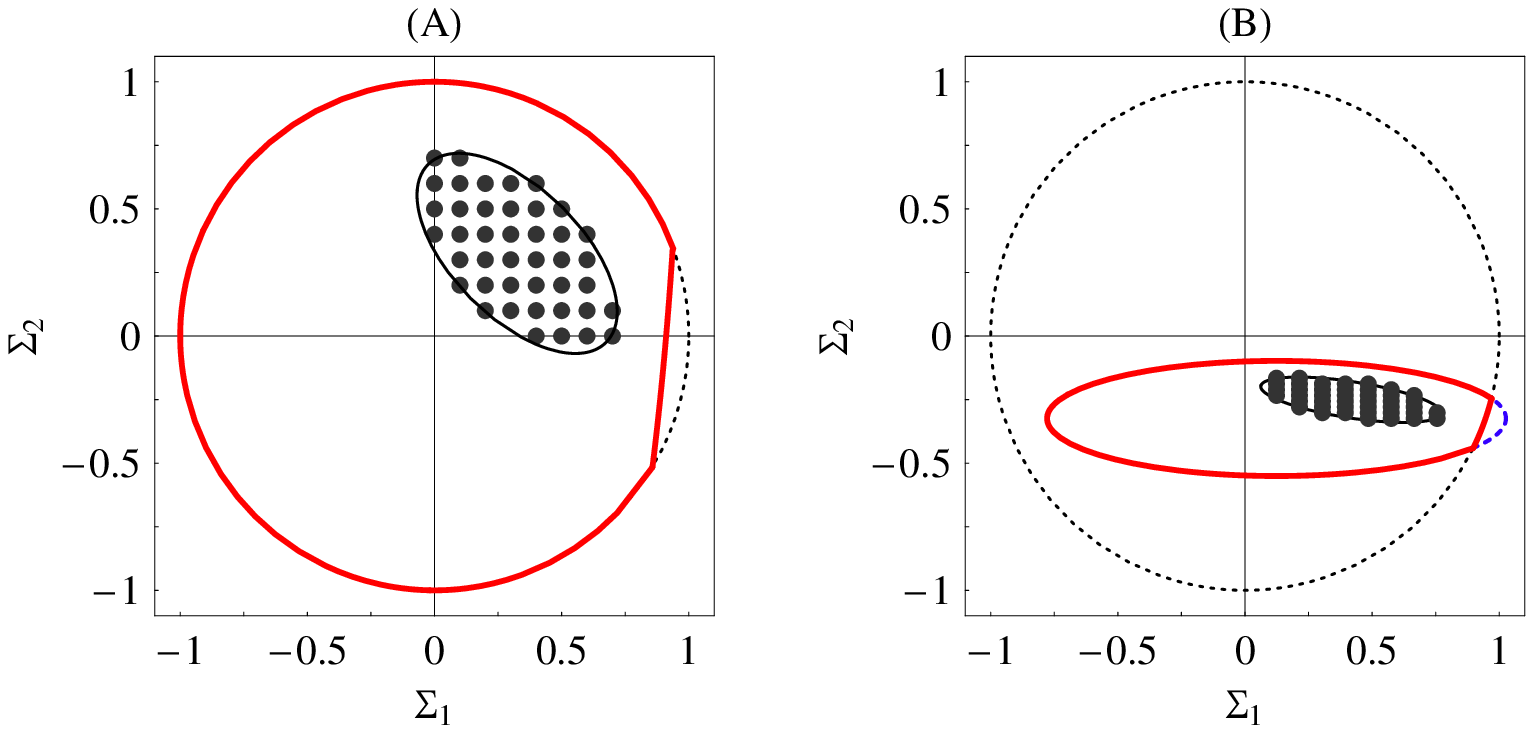}
   \end{center}
   \caption{(color online) (A) Sections in the $(\psig 1 \,,\, \psig 2)$ plane of the compatibility domain (dotted region) and the positivity domain (thick red curve)  for $\apxi{1}$, $\apxi{2}$, $\apxi{3}$, $\apsigxi{1}{2}$, $\apsigxi{1}{3}$, $\apsigxi{2}{1}$, $\apsigxi{2}{3}$, $\apsigxi{3}{1}$ and $\apsigxi{3}{2}$ all equal to $\frac{1}{4}$ and $\gamma_1=2 \sqrt{5}$, $\gamma_2= 2 \sqrt{3}$, $\gamma_3= 2 \sqrt{2}$. (B) Sections of where the map takes the compatibility domain, the positivity domain and the unit sphere (thick red curve plus the dashed blue curve). The dotted circle is the section of the unit sphere}
\label{fig1a}
\end{figure}

\begin{figure}[htb]
   \begin{center}
   \includegraphics{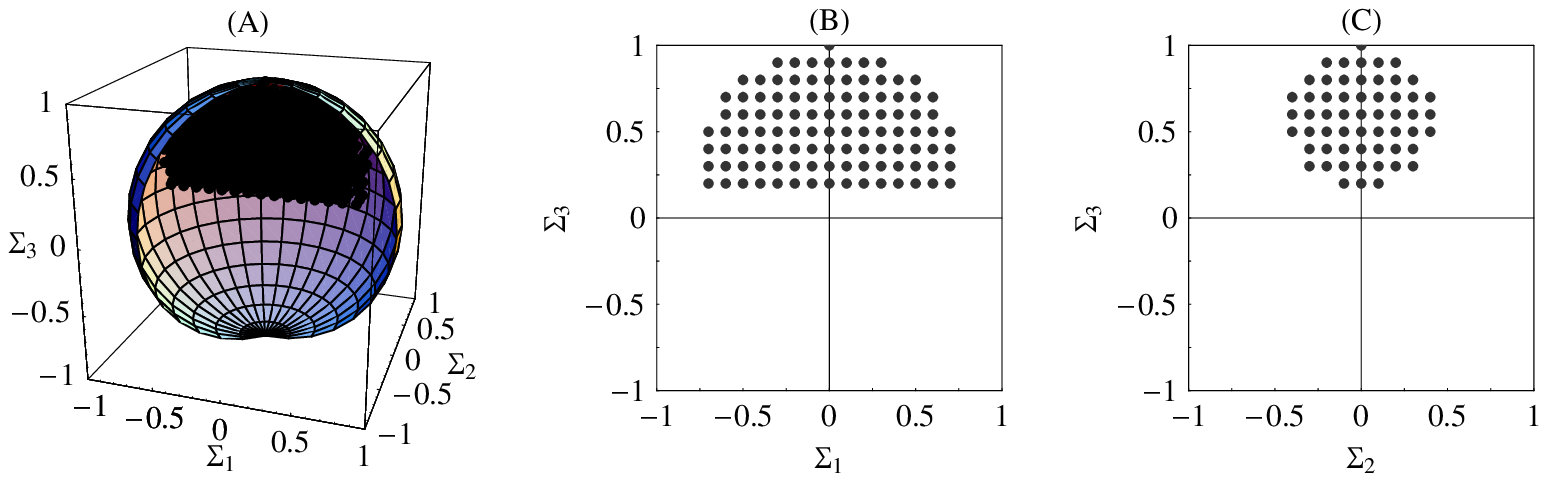}
   \end{center}
   \caption{(color online) The compatibility domain for $\apxi{1}$ and $\apsigxi{3}{1}$ equal to $\frac{1}{\sqrt{3}}$ and $\apxi{2}$, $\apxi{3}$, $\apsigxi{1}{2}$, $\apsigxi{1}{3}$, $\apsigxi{2}{1}$, $\apsigxi{2}{3}$,  $\apsigxi{3}{2}$ all equal to zero. (A) The whole compatibility domain inside the unit sphere. (B) The section of the domain in the $(\psig 1 , \psig 3)$ plane. (C) The section of the domain in the $(\psig 2 , \psig 3)$ plane. The compatibility domain does not intersect the $(\psig 1 , \psig 2)$ plane.}
\label{fig2}
\end{figure}
\end{widetext}

\noindent is positive (which implies that it is a density matrix and the set of $\avecsig$ is in the compatibility domain). Let
\begin{equation}
  \label{eq:int12}
  \Pi = \frac{1}{4} \left( \openone + \apxi{1} \pxi 1 + M \right).
\end{equation}
We see that $M$ commutes with $\pxi 1$. There is a basis of eigenvectors of $\pxi 1$ and $M$ that diagonalizes $\Pi$. From 
\begin{equation}
  \label{eq:int13}
  M^2 = \apsig{1}^2 + \apsig{2}^2 + \apsig{3}^2 + \apsigxi{3}{1}^2 + 2 \apsig{3} \apsigxi{3}{1} \pxi 1
\end{equation}
we see that the magnitude of the eigenvalues of $M$ is
\begin{equation}
  \label{eq:int14}
  \sqrt{\apsig{1}^2 + \apsig{2}^2 + \left( \apsig{3} \pm \apsigxi{3}{1} \right)^2}
\end{equation}
when the eigenvalue of $\pxi 1$ is $\pm 1$. The eigenvalues of $\Pi$ are all non-negative if
\begin{equation}
  \label{eq:int15}
  \apsig{1}^2 + \apsig{2}^2 + \left( \apsig{3} \pm \apsigxi{3}{1} \right)^2 \leq (1 \pm \apxi{1} )^2. 
\end{equation}
These inequalities say that $\avecsig$ is in the intersection of the two spheres of radii $1 \pm \apxi{1}$ with centers at $\mp \apsigxi{3}{1}$ on the $3$ axis. There is a substantial intersection, so there is a substantial compatibility domain, for all values of $\apxi{1}$ and $\apsigxi{3}{1}$ short of the limit where $\apxi{1}$ or $\apsigxi{3}{1}$ is 1. For example when $\apxi{1}$ and $\apsigxi{3}{1}$ are equal, the intersection is just the smaller sphere. 

Now we show that the zero $\avecsig$ is not in the compatibility domain when $\apxi{1}+\apsigxi{3}{1}$ is larger than 1. We show that then there is no density matrix 
\begin{eqnarray}
  \label{eq:int16}
  \Pi \! \!&=& \!\! \frac{1}{4} \left( \openone + \apsigxi{1}{1} \psig 1 \pxi 1 +  \apsigxi{2}{2} \psig 2 \pxi 2 +  \apsigxi{3}{3} \psig 3 \pxi 3 \right. \nonumber 
  \\ && \hspace{2 cm} \left. +   \apsigxi{3}{1} \psig 3 \pxi 1 + \apxi{1} \pxi{1} \right)
\end{eqnarray}
because then no matrix of this form can be positive. Let
\begin{equation}
  \label{eq:int17}
   \Pi = \frac{1}{4} \left( \openone  +  \apsigxi{2}{2} \psig 2 \pxi 2 +  \apsigxi{3}{3} \psig 3 \pxi 3  + \apxi{1} \pxi{1} + W \right)
\end{equation}
where
\begin{equation}
  \label{eq:int18}
  W = \left[  \apsigxi{1}{1} \psig 1 +   \apsigxi{3}{1} \psig 3 \right]\pxi 1.
\end{equation}
Because $W^2$ is $\apsigxi{1}{1}^2 +   \apsigxi{3}{1}^2$, the eigenvalues of $W$ are
\begin{equation}
  \label{eq:int19}
  \pm \sqrt{\apsigxi{1}{1}^2 +   \apsigxi{3}{1}^2}.
\end{equation}
The matrices $\pxi 1$ and $W$ commute and make a complete set of commuting $4$$\times$$4$ matrices. Their eigenvalues label basis vectors for the four-dimensional space, one for each of the four combinations of the two eigenvalues of $\pxi 1$ and the two eigenvalues of $W$. In particular, there is a nonzero vector $\Psi$ that is an eigenvector of $\pxi 1$ and $W$ for the negative eigenvalues of both. Since $(\Psi, \, \psig 2 \pxi 2 \Psi)$ and $(\Psi, \, \psig 3 \pxi 3 \Psi )$ are zero, it gives
\begin{equation}
  \label{eq:int20}
  (\Psi, \, \Pi \Psi) = \frac{1}{4}||\Psi||^2 \left( 1 - \apxi{1} - \sqrt{\apsigxi{1}{1}^2 +   \apsigxi{3}{1}^2} \right)
\end{equation}
which is negative if $\apxi{1} + \apsigxi{3}{1}$ is larger than 1. 

\subsection{Lorentz transformations of spin} \label{lorentz}

These examples are abstracted from Lorentz transformations of the spin of a particle with positive mass and spin $\frac{1}{2}$ for two possible values of the momentum \cite{jordan05a}. Let
\begin{equation}
  \label{eq:lt1}
  U = D_1 \frac{1}{2} \left( \openone + \pxi 1 \right) + D_2 \frac{1}{2} \left(  \openone - \pxi 1 \right)
\end{equation}
with $D_1$ and $D_2$ the unitary rotation matrices made from $\vecsig$, so that
\begin{equation}
  \label{eq:lt2}
  D_1^{\dagger} \vecsig D_1 = R_1(\vecsig) \,, \quad D_2^{\dagger} \vecsig D_2 = R_2 (\vecsig),
\end{equation}
for rotations $R_1$ and $R_2$; each $R(\vecsig)$ is simply the three-dimensional vector $\vecsig$ rotated by $R$. In the application to Lorentz transformations of spin \cite{jordan05a}, $\vecsig$ describes the spin of the particle, $\pxi 1$ is the Pauli matrix for states with two different momentum values ${\bf p}_1$ and ${\bf p}_2$ that is $+1$ when the momentum is ${\bf p}_1$ and $-1$ when the momentum is ${\bf p}_2$, and $R_1$ and $R_2$ are the Wigner rotations for the Lorentz transformation for ${\bf p}_1$ and ${\bf p}_2$. Then $U$ describes the Lorentz transformation of the spin in the system of two qubits where one qubit is the spin and the other is made from the two values of the momentum \cite{jordan05a}. We would not have thought to consider an example this simple had it not come to us in an interesting application.

From Eqs. (\ref{eq:lt1}) and (\ref{eq:lt2}) we get 
\begin{eqnarray}
  \label{eq:lt3}
  U^{\dagger} \vecsig U & = & R_1 (\vecsig) \frac{1}{2} \left( \openone + \pxi 1 \right) + R_2 (\vecsig) \frac{1}{2} \left( \openone - \pxi 1 \right) \nonumber \\
  & = & \frac{1}{2} \left[ R_1 (\vecsig) + R_2 (\vecsig) \right] + \frac{1}{2} \left[ R_1 (\vecsig) - R_2 (\vecsig) \right] \pxi 1, \nonumber \\
\end{eqnarray}
\begin{equation}
  \label{eq:lt4}
  L(\vecsig) =  \frac{1}{2} \left[ R_1^{-1} (\vecsig) + R_2^{-1} (\vecsig) \right],
\end{equation}
\begin{equation}
  \label{eq:lt5}
  \vec{\kappa} = \frac{1}{2} \left\langle \left[ R_1 (\vecsig) - R_2 (\vecsig) \right] \pxi 1 \right\rangle
\end{equation}
\begin{equation}
  \label{eq:lt6}
  G(0) = \frac{1}{2} \left( D_1 + D_2 \right) \,, \quad G(1) = \frac{1}{2} \left( D_1 - D_2 \right).
\end{equation}
The mean values $\avecsig$ are mapped to
\begin{equation}
  \label{eq:lt7}
  \avecsig^U = \langle U^{\dagger} \vecsig U \rangle = \frac{1}{2} \langle R_1 (\vecsig) + R_2 (\vecsig) \rangle + \vec{\kappa}
\end{equation}
and the density matrix $\rho$ of Eq. (\ref{eq:tq2}) is mapped to
\begin{eqnarray}
  \label{lt8}
  \rho^U & = & \frac{1}{2} \left( \openone + \avecsig^U \cdot \vecsig \right) \nonumber \\
  & = & \frac{1}{2} \left( \openone + \frac{1}{2}  \langle R_1 (\vecsig) + R_2 (\vecsig) \rangle \cdot \vecsig + \vec{\kappa} \cdot \vecsig \right) \nonumber \\
  & = & \frac{1}{2} \left( \openone + \avecsig \cdot \frac{1}{2} \left[ R_1^{-1} (\vecsig) + R_2^{-1} (\vecsig) \right] + \vec{\kappa} \cdot \vecsig \right) \nonumber \\
  & = & L(\rho) + K  \nonumber \\
  &=& \frac{1}{2} \left( \openone' + \avecsig \cdot \vecsig' \right)
\end{eqnarray}
with
\begin{equation}
  \label{eq:lt9}
  \vecsig' = L (\vecsig) \,, \quad \openone' = \openone + 2 K.
\end{equation}

If $\langle R_1 (\vecsig) \pxi 1 \rangle = \langle R_2 (\vecsig) \pxi 1 \rangle$ then $K$ is zero and the map is completely positive. In fact the map is the same as it would be if $\langle \vecsig \pxi 1 \rangle$ were zero. If $\langle \vecsig \pxi 1 \rangle$ is zero then
\begin{equation}
  \label{eq:lt10}
  \left\langle \vecsig \frac{1}{2} ( \openone + \pxi 1) \right\rangle = \left\langle \vecsig \frac{1}{2} ( \openone - \pxi 1 ) \right\rangle. 
\end{equation}
In the application \cite{jordan05a}, the mean value of the spin is the same for both momenta ${\bf p}_1$ and ${\bf p}_2$. 

If $\langle R_1 (\vecsig) \pxi 1 \rangle \neq \langle R_2 (\vecsig) \pxi 1 \rangle$ there are positive matrices
\begin{equation}
  \label{eq:lt11}
  \frac{1}{2}( \openone + \vec{a} \cdot \vecsig )
\end{equation}
that are mapped to matrices
\begin{equation}
  \label{eq:lt12}
  \frac{1}{2} ( \openone + \vec{a}' \cdot \vecsig ) 
\end{equation}
that are not positive. To see this it is sufficient to consider the case where $R_1$ is the identity rotation and $R_2$ is an arbitrary rotation $R$. The general case can be recovered by taking $R$ to be $R_1^{-1}R_2$ and joining the same rotation $R_1$ onto both the identity and $R_1^{-1}R_2$ to restore $R_1$ and $R_2$. The overall rotation $R_1$ will just rotate $\vec{a}'$ and not change the non-positive character of the matrix (\ref{eq:lt12}). Hence we consider
\begin{equation}
  \label{eq:lt13}
  \vec{a}' = \frac{1}{2} \left[ \vec{a} + R(\vec{a}) \right] + \frac{1}{2} \left\langle \left[ \vecsig - R(\vecsig) \right] \pxi 1 \right\rangle.
\end{equation}
Let $\vec{a}$ be along the axis of $R$ so that $R(\vec{a})$ is $\vec{a}$. Then
\begin{equation}
  \label{eq:lt14}
  \vec{a}' = \vec{a} + \frac{1}{2} \left\langle \left[ \vecsig - R(\vecsig) \right] \pxi 1 \right\rangle.
\end{equation}
Choose the direction of $\vec{a}$ so that
\begin{equation}
  \label{eq:lt15}
  \vec{a} \cdot \frac{1}{2} \left\langle \left[ \vecsig - R(\vecsig) \right] \pxi 1 \right\rangle \geq 0.
\end{equation}
Then $|\vec{a}'| > |\vec{a}|$. When $|\vec{a}|$ approaches 1, the matrix (\ref{eq:lt12}) is not positive.  

For these maps, $K$ depends only on $\langle \vecsig \pxi 1 \rangle$, not on $\langle \vecsig \pxi 2 \rangle$, $\langle \vecsig \pxi 3 \rangle$ or $\avecxi$. The compatibility domain is the set of $\avecsig$ that are compatible with specified $\langle \vecsig \pxi 1 \rangle$ in describing a possible state for the two qubits. It is very similar to the compatibility domain for the examples described in our previous paper. More precisely, $K$ depends on
\begin{equation}
  \label{eq:lt16}
  R_1 (\langle \vecsig \pxi 1 \rangle)-R_2  (\langle \vecsig \pxi 1 \rangle) = R_1 \left( \langle \vecsig \pxi 1 \rangle - R_1^{-1}R_2(\langle \vecsig \pxi 1 \rangle) \right). 
\end{equation}
Let the $3$ axis be along the axis of $ R_1^{-1}R_2$. Then $\apsigxi{3}{1}$ is not changed by $ R_1^{-1}R_2$, so it drops out, leaving only $\apsigxi{1}{1}$ and $\apsigxi{2}{1}$ in $K$, and the compatibility domain is the set of $\avecsig$ that are compatible with specified $\apsigxi{1}{1}$ and $\apsigxi{2}{1}$ in describing a possible state for the two qubits. This is exactly the compatibility domain for the examples described in our previous paper \cite{jordan04}. The equations and drawings that describe the compatibility domain there (Eqs. (2.58), (2.64), (2.75), (2.77) and \linebreak Figs. 2 and 3 in \cite{jordan04}) apply here as well.

The positivity domain is not the same as for the examples described in our previous paper \cite{jordan04}. It is the domain in which every positive matrix is mapped to a positive matrix, or the set of $\avecsig$ for which
\begin{equation}
  \label{eq:lt17}
  \left| \frac{1}{2} \left[ R_1 (\avecsig) + R_2 (\avecsig) \right] + \vec{\kappa} \right| = \left| \avecsig^U \right| \leq 1. 
\end{equation}

\subsection{The size of the inhomogeneous part} \label{size}

How big can $|\vec{\kappa}|$ be? For the Lorentz-transformation examples described in Section \ref{lorentz} we find that the limit on $|\vec{\kappa}|$ is 1, but that $|\vec{\kappa}|$ can have any value short of that limit \cite{jordan05a}. We do not know if values of $|\vec{\kappa}|$ larger than 1 are possible for the interaction-Hamiltonian examples described in Section \ref{interaction}; we have not found any values larger than 1. Here is a proof that $|\vec{\kappa}|$ cannot be larger than $\frac{1+\sqrt{5}}{2}$, which is about 1.62. 

From Eqs. (\ref{eq:tq4}) and (\ref{eq:a4}) we have
\begin{eqnarray}
  \label{eq:si1}
  \kappa_j &=& \Tr_S \left[ \psig j \Tr_R \left[ U \left( \Pi - \rho \otimes \frac{\openone}{2} \right) U^{\dagger} \right] \right] \nonumber \\
  &=& \Tr \left[ U^{\dagger} \psig j U \left( \Pi - \rho \otimes \frac{\openone}{2} \right) \right]
\end{eqnarray}
with $\Pi$ and $\rho$ the density matrices of Eqs. (\ref{eq:tq1}) and (\ref{eq:tq2}). Since the trace of the product of matrices has the properties of an inner product for the real linear space of Hermitian matrices, it follows that
\begin{eqnarray}
  \label{eq:si2}
  (\kappa_3)^2 \!\! &\leq& \!\!\! \Tr \left[ \left( U^{\dagger} \psig 3 U \right)^2 \right] \Tr \left[ \left( \Pi - \rho \otimes \frac{\openone}{2} \right)^2 \right] \:\:\:\nonumber \\
  \!\!&=& \!\!\!4 \left( \frac{1}{4} \right) \!\!\! \left( \sum_{k=1}^3 \apxi{k}^2 + \sum_{j,k=1}^3 \apsigxi{j}{k}^2 \right)\!. \;\;\; 
\end{eqnarray}
From 
\begin{eqnarray}
  \label{eq:si3}
  1 & \geq & \Tr \left[ \Pi^2 \right] \nonumber \\
  & = & \frac{1}{4} \left( \openone + \sum_{j=1}^3 \apsig{j}^2 + \sum_{k=1}^3 \apxi{k}^2 + \sum_{j,k=1}^3 \apsigxi{j}{k}^2 \right) \nonumber \\
\end{eqnarray}
we get
\begin{equation}
  \label{eq:si4}
  (\kappa_3)^2 \leq 3 - \sum_{j=1}^3 \apsig{j}^2.
\end{equation}
This holds for any unitary matrix $U$, so it holds when $U$ is replaced by $UV$ with $V$ a unitary rotation matrix made from the Pauli matrices $U^{\dagger} \psig j U$ that rotates $U^{\dagger} \vecsig U$ to $V^{\dagger}U^{\dagger} \vecsig UV$ so that the only nonzero component of $\vec{\kappa}$ for $V^{\dagger}U^{\dagger} \vecsig UV$ is $\kappa_3$ and $|\vec{\kappa}|^2$ for $U$ is $(\kappa_3)^2$ for $UV$. Thus we conclude that
\begin{equation}
  \label{eq:si4a}
  |\vec{\kappa}|^2 \leq 3 - |\avecsig|^2.
\end{equation}

We consider mean values $\apsig{j}$, $\apxi{k}$, $\apsigxi{j}{k}$ that describe a possible initial state for the two qubits. Then also
\begin{eqnarray}
  \label{eq:si5}
  |\vec{\kappa}| &=& \left| \Tr_S \left[ \vecsig K \right] \right| \nonumber \\
  &=& \left| \left\langle U^{\dagger} \vecsig U \right \rangle - \Tr_S \left[ \vecsig L (\rho) \right] \right| \leq 1 + \left| \avecsig \right| \nonumber \\
\end{eqnarray}
because $ \left| \left\langle U^{\dagger} \vecsig U \right \rangle \right| \leq 1$ and, as in Eq. (\ref{eq:p7})
\begin{eqnarray}
  \label{eq:si6}
  \sum_{j=1}^3 \left( \Tr_S \left[ \psig j L (\rho) \right] \right)^2 &=& \sum_{j=1}^3 \left( \langle F_{j0} \rangle_0^U \right)^2 \nonumber \\
  &&\leq \sum_{j=1}^3 \langle F_{j0} \rangle^2 = \sum_{j=1}^3 \apsig{j}^2. \nonumber \\
\end{eqnarray}

As a function of $| \avecsig |$, the bound (\ref{eq:si4a}) decreases and the bound (\ref{eq:si5}) increases. The two bounds allow the largest $| \vec{\kappa}|$ when they meet. Then $| \avecsig |$ is $\frac{-1 + \sqrt{5}}{2}$ and the largest $| \vec{\kappa}|$ allowed is $\frac{1 + \sqrt{5}}{2}$. As the limit of large $|\vec{\kappa}|$ is approached, the room for variation in $|\avecsig|$ decreases, so room for the compatibility domain decreases, as our examples have shown.

\bibliography{ncp}
\end{document}